\documentclass[preprint,12pt]{elsarticle}

\usepackage{epsfig}

\usepackage{amssymb}
\usepackage{url}
\usepackage{graphicx}
\usepackage{color}

\usepackage{amssymb}
\usepackage{amsmath}
\usepackage{latexsym}
\usepackage{enumerate}
\usepackage{multirow}
\usepackage{twoopt}
\usepackage{array}
\usepackage{alltt}

\usepackage[nodots,nocompress]{numcompress}

\biboptions{super}

\journal{Physica A: Statistical Mechanics and its Applications}

\begin{document}

\begin{frontmatter}



\title{Social Influences in Opinion Dynamics: the Role of Conformity}

\author[a,b]{Marco Alberto Javarone}
\ead{marcojavarone@gmail.com}
\address[a]{DUMAS - Dept. of Humanities and Social Sciences - University of Sassari - 07100, Sassari, Italy}
\address[b]{Dept. of Mathematics and Computer Science - University of Cagliari - P.zza D'armi 09123
Cagliari, Italy}

\begin{abstract}
We study the effects of social influences in opinion dynamics. In particular, we define a simple model, based on the majority rule voting, in order to consider the role of conformity. Conformity is a central issue in social psychology as it represents one of people’s behaviors that emerges as a result of their interactions. The proposed model represents agents, arranged in a network and provided with an individual behavior, that change opinion in function of those of their neighbors. In particular, agents can behave as conformists or as nonconformists. In the former case, agents change opinion in accordance with the majority of their social circle (i.e., their neighbors); in the latter case, they do the opposite, i.e., they take the minority opinion. Moreover, we investigate the nonconformity both on a global and on a local perspective, i.e., in relation to the whole population and to the social circle of each nonconformist agent, respectively. We perform a computational study of the proposed model, with the aim to observe if and how the conformity affects the related outcomes. Moreover, we want to investigate whether it is possible to achieve some kind of equilibrium, or of order, during the evolution of the system. Results highlight that the amount of nonconformist agents in the population plays a central role in these dynamics. In particular, conformist agents play the role of stabilizers in fully-connected networks, whereas the opposite happens in complex networks. Furthermore, by analyzing complex topologies of the agent network, we found that in the presence of radical nonconformist agents the topology of the system has a prominent role; otherwise it does not matter since we observed that a conformist behavior is almost always more convenient. Finally, we analyze the results of the model by considering that agents can change also their behavior over time, i.e., conformists can become nonconformists and vice versa.
\end{abstract}

\end{frontmatter}

%\tableofcontents

\section{Introduction}\label{introduction}
During last years opinion dynamics~\cite{loreto01}\cite{sznajd01} have been investigated by several authors. Just to cite a few, Galam introduced many sociophysics models~\cite{galam01}\cite{galam02}\cite{galam03} as, for instance, a spin model for studying the dynamics of majority rules~\cite{galam04} that, later on, has been further developed by Krapivsky and Redner~\cite{krapviski01}, Holme and Newman analyzed the opinion spreading in terms of a nonequilibrium phase transition~\cite{holme01}, and Bianconi and collaborators proposed a model to consider the role of social networks in these dynamics~\cite{bianconi01}.
The voter model~\cite{redner01} represents one of the most famous models of opinion dynamics.
Usually, this model considers a set of interacting agents, provided with a state that represents their opinion. In so doing, it is possible to perform computational studies to analyze the evolution toward consensus in the presence of different opinions.
Agent-based models allow to study interesting phenomena in opinion dynamics (see for instance~\cite{miguel01}\cite{miguel02}\cite{eger01}) or, more in general, in social dynamics~\cite{rapisarda01}\cite{rapisarda02}\cite{williams01}.
We put our attention on the majority rule voting~\cite{galam03} from a socio-psychological perspective~\cite{aronson01}. Social psychology provides fundamental theories to study interactions among individuals in social contexts, for instance see~\cite{javarone01}. 
In this work, we analyze the role of conformity, an important behavior of individuals~\cite{aronson01} that emerges as result of their interactions, by a simple model based on the majority rule voting.
In the proposed model, we consider a system with only two opinions (i.e., two possible states) and we provide agents with an individual behavior; in particular, they can be conformists or nonconformists (i.e., contrarians). 
Conformist agents take the opinion of the majority of their neighbors, whereas the nonconformist agents do the opposite.
As discussed in~\cite{galam05}, the contrarian effects have an important role in the voter model (see also~\cite{weron01}\cite{weron02}\cite{buechel01}) and they can also explain complex phenomena in real electoral dynamics.
We analyze the nonconformity under two different hypotheses, i.e., a local nonconformity and a global nonconformity. The former is a behavior related only to the social circle of each agent, but not to the whole population. 
This concept means that a nonconformist agent takes the minority opinion of its social circle (i.e., its neighbors), but it prefers to have the same opinion of majority of the population. Instead, the global nonconformity is more radical, as it assumes that nonconformist agents aim to have an opinion contrarian also to that of the majority of the population.
We perform a computational study of the proposed model by considering different conditions, as the topology of the agent network and the density of nonconformist agents in the population.
Moreover, we analyze the system by allowing agents to change behavior over time. In particular, at each time step, agents decide whether to be conformists or nonconformists, depending on the comparison between their state and that of the majority of the population, and to their current behavior.
The main result is that conformity is an important behavior in these dynamics as it strongly affects the outcomes of the proposed model. Furthermore, we found important differences between the outcomes achieved by varying the topology of the agent network if a radical nonconformity is considered; otherwise, with different topologies, the outcomes are always very similar.
The remainder of the paper is organized as follows: Section~\ref{sec:model} introduces the model for studying the role of conformity. Section~\ref{sec:results} shows results of numerical simulations. Eventually, Section~\ref{sec:conclusions} ends the paper.
\section{Conformists versus Non-Conformists}\label{sec:model}
\begin{figure}[!ht]
\centering
\includegraphics[width=3in]{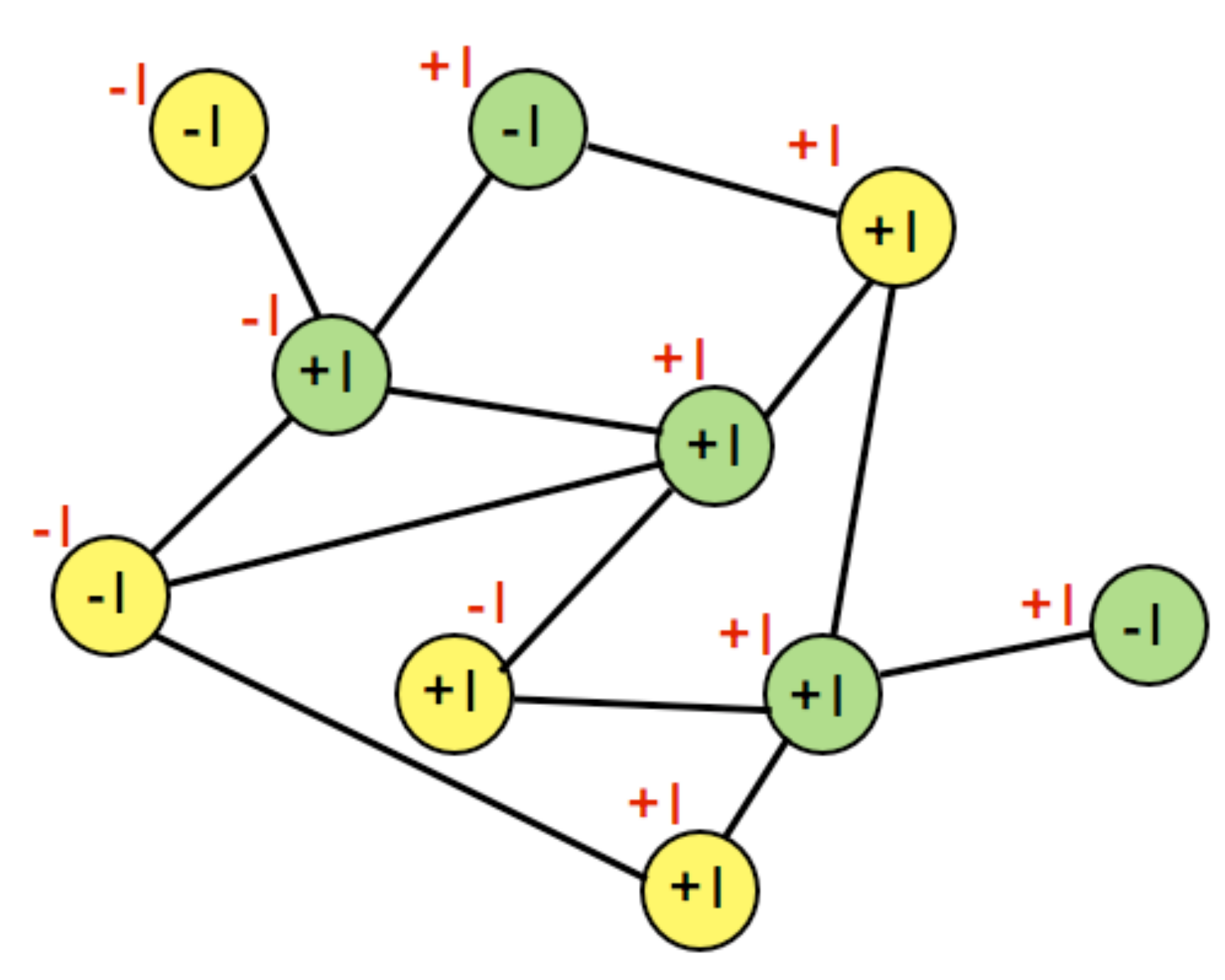}
\caption{A small network of conformist agents (i.e., green nodes) and nonconformist agents (i.e., yellow nodes). The inner numbers represent the state of agents at time $t$, whereas the red numbers represent the state of agents at time $t+1$.}
\label{fig:small_network}
\end{figure}

We introduce a simple model of opinion dynamics where agents interact over a network. Agents have an opinion, mapped to a state $s= \pm 1$ (i.e., there are two possible opinions) and, in addition, they are provided with an individual behavior. In particular, an agent can be conformist or nonconformist. 
Conformist agents modify their state (i.e., opinion) according to the majority of their neighbors, whereas nonconformist ones do the opposite, i.e., they take the minority opinion.
Therefore, conformist agents change state over time as follows:
\begin{equation} \label{eq:conformist_0}
s_{i}(t+1) = \begin{cases} +1 & \mbox{if } \sum_{j=1}^{n^{i}} s_j(t) >\ 0 \\ 
-1 & \mbox{if } \sum_{j=1}^{n^{i}} s_j(t) <\ 0 \\
s_{i}(t) & \mbox{if } \sum_{j=1}^{n^{i}} s_j(t)  = 0 \end{cases}
\end{equation}
\noindent with $s_i(t)$ state of the $i$th agent at time $t$, $n^{i}$ number of neighbors of the $i$th agent and $s_j(t)$ state of the $j$th agent, linked with the $i$th agent.
On the other hand, nonconformist agents follow an opposite rule for changing their state:
\begin{equation} \label{eq:non-conformist_0}
s_{i}(t+1) = \begin{cases} +1 & \mbox{if } \sum_{j=1}^{n^{i}} s_j(t) <\ 0 \\ 
-1 & \mbox{if } \sum_{j=1}^{n^{i}} s_j(t) >\ 0 \\
s_{i}(t) & \mbox{if } \sum_{j=1}^{n^{i}} s_j(t)  = 0 \end{cases}
\end{equation}
In so doing, at each time step, agents compute the new state depending on their behavior and on the opinion of their social circle. 
Figure~\ref{fig:small_network} shows an example of the proposed model.
Furthermore, agents achieve a score computed by comparing their state $s(t)$ with that of the whole system $S(t)$, defined as $S(t) = \sum_{i=1}^{N}s_i(t)$ (i.e., the total sum of states).
In particular, we define two different kinds of scores: the local score $P$ and the global scores $(P^{c}$, $P^{a})$. 
These scores allow to consider two different hypotheses:
\begin{itemize}
\item [\textbf{a})] The nonconformity is related only to the social circle of each (nonconformist) agent, but all agents prefer to have a state in accordance with that of majority of the whole population (i.e., $S(t)$);
\item [\textbf{b})] The nonconformity is related to the whole population, therefore nonconformist agents aim to get a state opposed to that of majority of the whole population.
\end{itemize}
The conceptual difference between these two hypotheses is that, using the first one we consider that the nonconformity behavior has a local dimension, whereas by the second one the nonconformity behavior is more radical as it is related to the whole population.
Therefore, under the hypothesis \textbf{a}, agents increase their local score $P$ every time their state is in accordance with the value of $S(t)$ (i.e., both values are positive or negative). 
For instance, in the event the $j$th agent has the state $s_j(t) = -1$ and the summation of states is equal to $S(t)= -120$, its $P_j$ increases of $+1$; otherwise it decreases (of $-1$) or, in the event $S(t)=0$, its local score does not change.
On the other hand, considering the hypothesis \textbf{b}, we adopt the global scores $(P^{c}$, $P^{a})$, with $P^{c}$ defined for nonconformist agents, and $P^{a}$ for conformist ones. 
The score $P^{a}$ increases when the state $s(t)$ of a conformist agent is in accordance with $S(t)$, whereas it decreases in the opposite case. On the contrary, the score $P^{c}$ increases when the state $s(t)$ of a nonconformist agent is opposite to $S(t)$ (and it decreases when it is in accordance with $S(t)$). 
A state $s(t)$ in accordance with that of the majority of the population means that $s(t)$ and $S(t)$ have the same sign, whereas on the contrary, an opposite state means that $s(t)$ and $S(t)$ have not the same sign.
As for the local score $P$, the increasing and the decreasing of global scores $(P^{c}, P^{a})$, at each time step, is equal to $+1$ and $-1$, respectively. Moreover $(P^{c}$, $P^{a})$ do not change if $S(t)=0$.
In general, as discussed before, the local score $P$ allows to evaluate whether it is more advantageous to behave as a conformist or as a nonconformist agent considering, as hypothesis, that the nonconformist behavior is related only the the social circle of each agent. 
Instead, the global scores $(P^{c}$, $P^{a})$ allow to evaluate whether nonconformist agents obtain benefits by having a more radical behavior in relation to the whole population.
Summarizing, the proposed model can be described as follows:
\begin{enumerate}
\item Define a network with $N$ agents provided with a random state $\pm 1$ and with a behavior (i.e., conformist or nonconformist);
\item At each time step, every $j$th agent computes the summation of states of its social circle ($S_j(t)$). Therefore, it changes its state $s_j(t)$ in function both of its behavior and of the value $S_j(t)$: in the event the $j$th agent is conformist, it changes its state in accordance with $S_j(t)$ whereas, if it is nonconformist, it changes its state to the opposite of $S_j(t)$. In both cases, if $S_j(t) =0$, the state of the $j$th agent does not change; 
\item The system evolves until a steady-state is reached (or a maximum number of time steps is elapsed).
\end{enumerate}
Finally, we introduce a small variation of the proposed model in order to let agents to change also their behavior over time. 
Also in this case, we consider the two hypotheses, i.e., \textit{a} and \textit{b}, about the nonconformist behavior.
Under the hypothesis \textit{a}, the following rule holds: at each time step, agents compare their state $s(t)$ with the total summation $S(t)$; in the event $s(t)$ and $S(t)$ have a different sign they change behavior.
Instead, considering the second hypothesis (i.e., \textit{b}), the rule to let agents change behavior is defined as follows: at each time step, a nonconformist agent changes behavior if its state is in accordance with $S(t)$, whereas it does not in the opposite case. 
On the other hand, a conformist agent becomes a nonconformist one if its state is not in accordance with $S(t)$, whereas its behavior does not change in the opposite circumstance.
In particular, a conformist agent becomes a nonconformist one, at time $t+1$, if its state $s(t)$ is not in accordance with $S(t)$, whereas a nonconformist agent becomes a conformist one (at time $t+1$) if its state $s(t)$ is in accordance with $S(t)$.
This variation of the model allows to observe if agents find more convenient to behave as conformists or as nonconformists over time, considering the two levels of nonconformity defined above (i.e., the two hypotheses \textit{a} and \textit{b}).
\section{Simulations}\label{sec:results}
We perform numerical simulations of the proposed model in a simple fully-connected agent network and, later on, we consider more complex topologies as scale-free networks and small-world networks.
Scale-free networks have been generated by the Barabasi-Albert model~\cite{barabasi01}, whereas small-world networks by the Watts-Strogatz model~\cite{watts01}. 
In particular, to generate small-world networks, we started from a $2$-dimensional regular lattice with $6$ neighbors per node, then we rewired with probability $\beta = 0.1$ each edge at random. 
At $t=0$, each agent has a state that can be $-1$ or $+1$ both with probability $0.5$ hence, on average, at the beginning the two opinions are equally assigned in the population.
The parameter of control is the density of conformist agents in the population $\rho_{a}$. Varying the value of $\rho_{a}$, it is possible to compare the outcomes of the proposed model achieved by a different initial condition (i.e., the value of $\rho_{a}$). In our hypothesis (and also according to theories of social-psychology), varying the value of $\rho_{a}$ the evolution of the system should be strongly affected.
Eventually, we perform simulations by allowing agents to change their behavior, i.e., from conformist to nonconformist, and vice versa. 
Therefore, in this last case, the parameter of control is $\rho_{a}(0)$, i.e., the density of conformist agents at time $t=0$.
\subsection{Fully-Connected Agent Network}
\begin{figure}[!ht]
\centering
\includegraphics[width=5.5in]{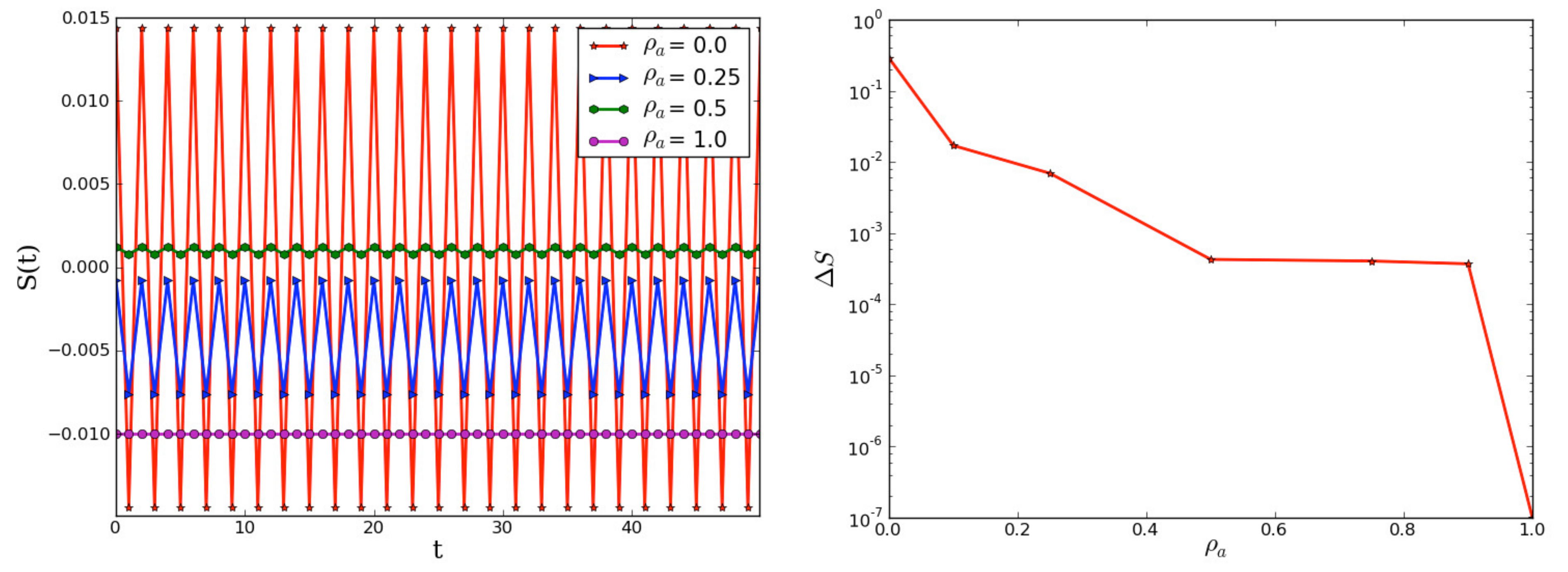}
\caption{On the left, the variation of $S(t)$ in fully-connected networks. On the right, the difference between the maximum and the minimum value of $S(t)$, indicated as $\Delta S$, varying $\rho_{a}$. Results are averaged over $20$ different realizations.}
\label{fig:sum_states_fc}
\end{figure}

\begin{figure}[!ht]
\centering
\includegraphics[width=5.5in]{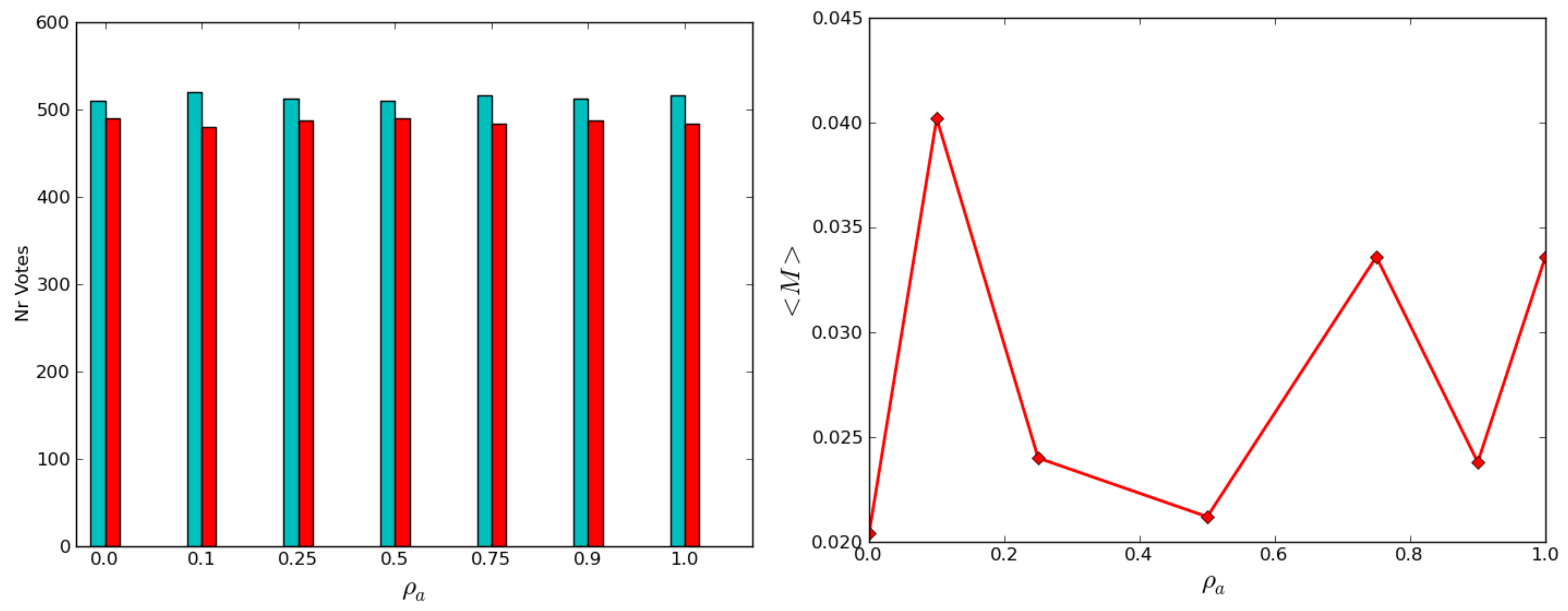}
\caption{On the left, number of agents in the two states, codified by two colors. The cyan represents the number of agents having the predominant opinion, whereas the red that of agents with the other opinion. On the right the average magnetization $<M>$ of the system in function of $\rho_{a}$. Results are averaged over $20$ different realizations.}
\label{fig:magnetization_fc}
\end{figure}

The first analysis is performed by using a fully-connected network with $N=1000$ agents. Figure~\ref{fig:sum_states_fc} shows the summation of states $S(t)$, in the whole population over time, varying $\rho_{a}$. In all cases, at each time step, the value of $S(t)$ oscillates between two values $S_{max}$ and $S_min$. 
It is interesting to note that, as $\rho_{a}$ increases the difference between $S_{max}$ and $S_min$ (i.e., $\Delta S$) decreases, falling almost to zero for $\rho_{a}=1$. 
Therefore, under this configuration of the system, the density of conformist agents strongly affects the outcomes of the proposed model.
In this kind of models, an important value to be analyzed is the average magnetization $<M>$ of the system, i.e., the diﬀerence in the density of agents in the two states~\cite{mobilia01} --see Figure~\ref{fig:magnetization_fc}.
In general, we found small differences between the number of agents with the two opinions. Hence, it seems that the value of $\rho_{a}$ does not affect the average magnetization of the system, i.e., there is not any relation between $\rho_{a}$ and $<M>$.
\subsection{Complex Agent Networks}
\begin{figure}[!ht]
\centering
\includegraphics[width=5.5in]{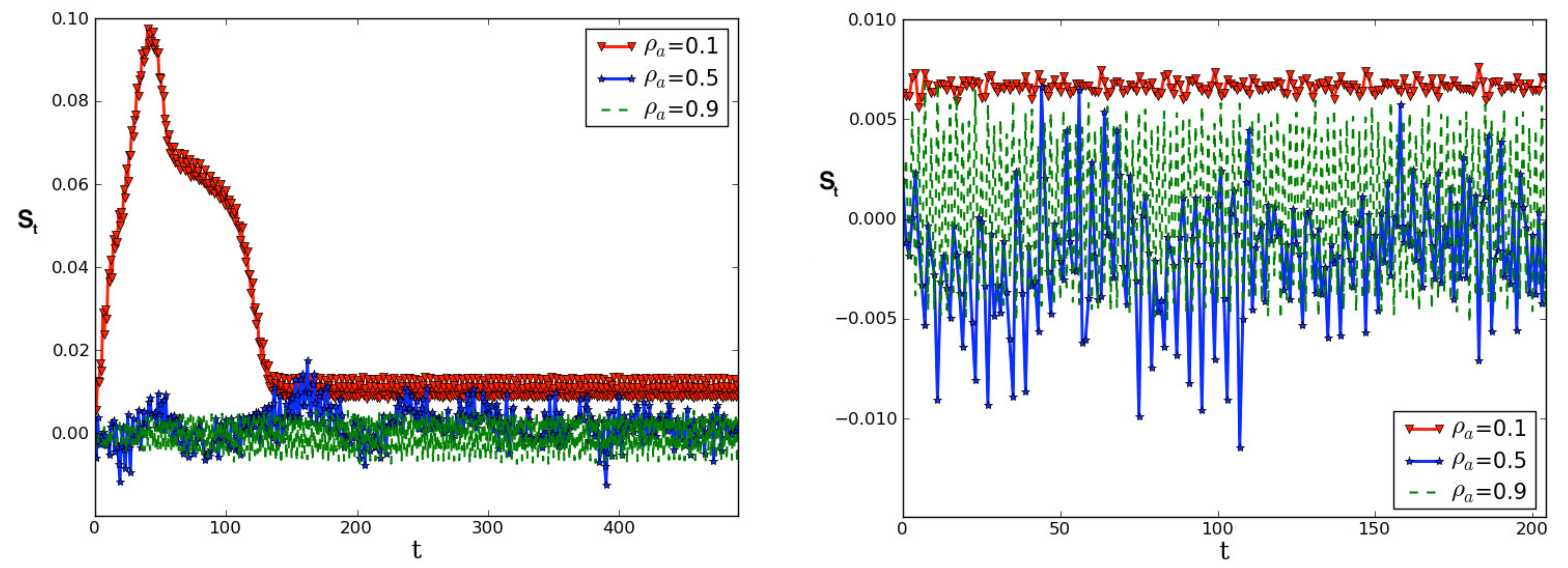}
\caption{On the left, the variation of $S(t)$ in scale-free networks. On the right, the variation of $S(t)$ in small-world networks. Each curve refers to results achieved varying $\rho_{a}$ as indicated in the legend. Results are averaged over $20$ different realizations.}
\label{fig:sum_states_complex}
\end{figure}

\begin{figure}[!ht]
\centering
\includegraphics[width=5.5in]{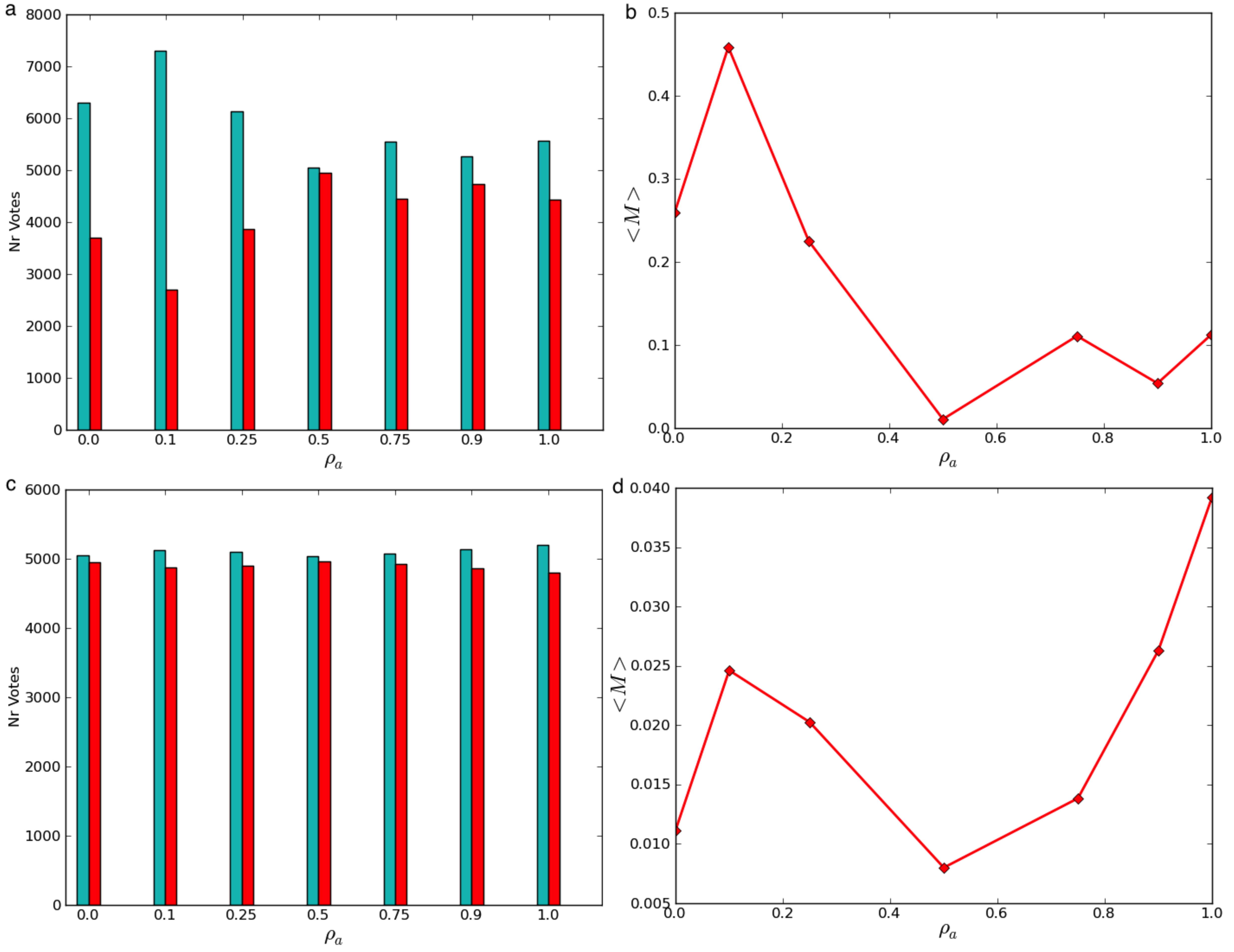}
\caption{\textbf{a} and \textbf{c} represent the number of agents in the two states (codified by two colors), in scale-free networks and in small-world networks, respectively. In particular, the cyan represents the number of agents having the predominant opinion, whereas the red that of agents with the other opinion. \textbf{b} and \textbf{d} represent the average magnetization of the system, in function of $\rho_{a}$, in scale-free networks and in small-world networks, respectively. Results are averaged over $20$ different realizations.}
\label{fig:magnetization_complex}
\end{figure}

As discussed above, we analyze the proposed model by using more complex topologies of the agent network, i.e., scale-free networks and small-world networks. 
Both kinds of complex networks have been generated with $N=10^4$ agents and with an average degree $\langle k\rangle = 6$.
As shown in Figure~\ref{fig:sum_states_complex}, the evolution of the system, in terms of the summation of states $S(t)$, is similar in both agent networks. In general, $S(t)$ fluctuates around an average value that is positive in the event there are a few conformist agents, whereas it fluctuates around zero (or around lightly negative values) as the amount of conformist agents (i.e., $\rho_{a}$) increases.
It is worth to highlight that, in scale-free networks, the $S(t)$ reaches a global maximum during first time steps for $\rho_{a} <\ 0.25$. In particular, $S(t)$ has a rapid increase, indicating that a great fraction of agents changes its state to $+1$, followed by a decrease of $S(t)$ up to a small positive average value.
As before, we analyze the average magnetization of the system in these two kinds of agent network --see Figure~\ref{fig:magnetization_complex}.
Although on a quality level, the values of the average magnetization are similar in both kinds of complex networks, they are very different quantitatively. In particular, as panels \textbf{a} and \textbf{c} of Figure~\ref{fig:magnetization_complex} show, there is a higher difference between agents having a different opinion in scale-free networks than in small-world networks.
\subsubsection*{Conformists vs Non-Conformists}
\begin{figure}[!ht]
\centering
\includegraphics[width=5.5in]{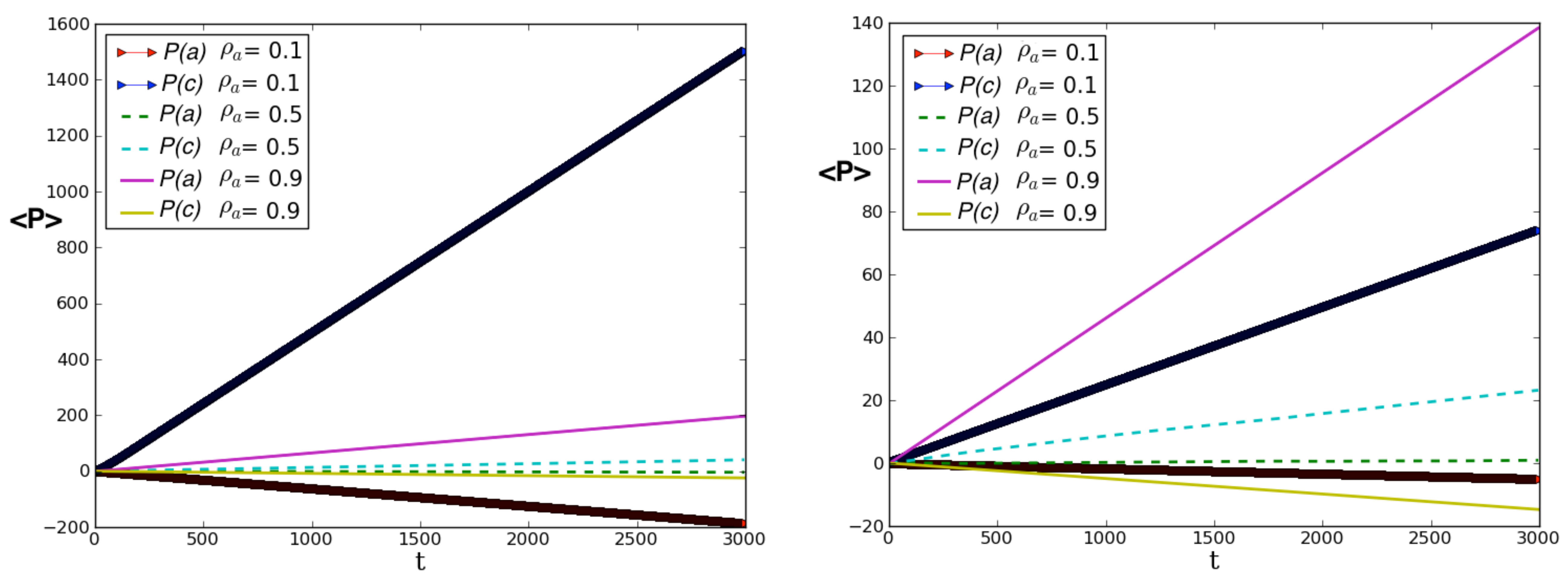}
\caption{Value of the average scores $\langle P \rangle$, over time, achieved by conformist agents ($P(c)$) and nonconformist agents ($P(a)$) varying $\rho_{a}$. On the left, results achieved in scale-free networks. On the right results achieved in small-world networks. Results are averaged over $20$ different realizations.}
\label{fig:score_P}
\end{figure}

\begin{figure}[!ht]
\centering
\includegraphics[width=5.5in]{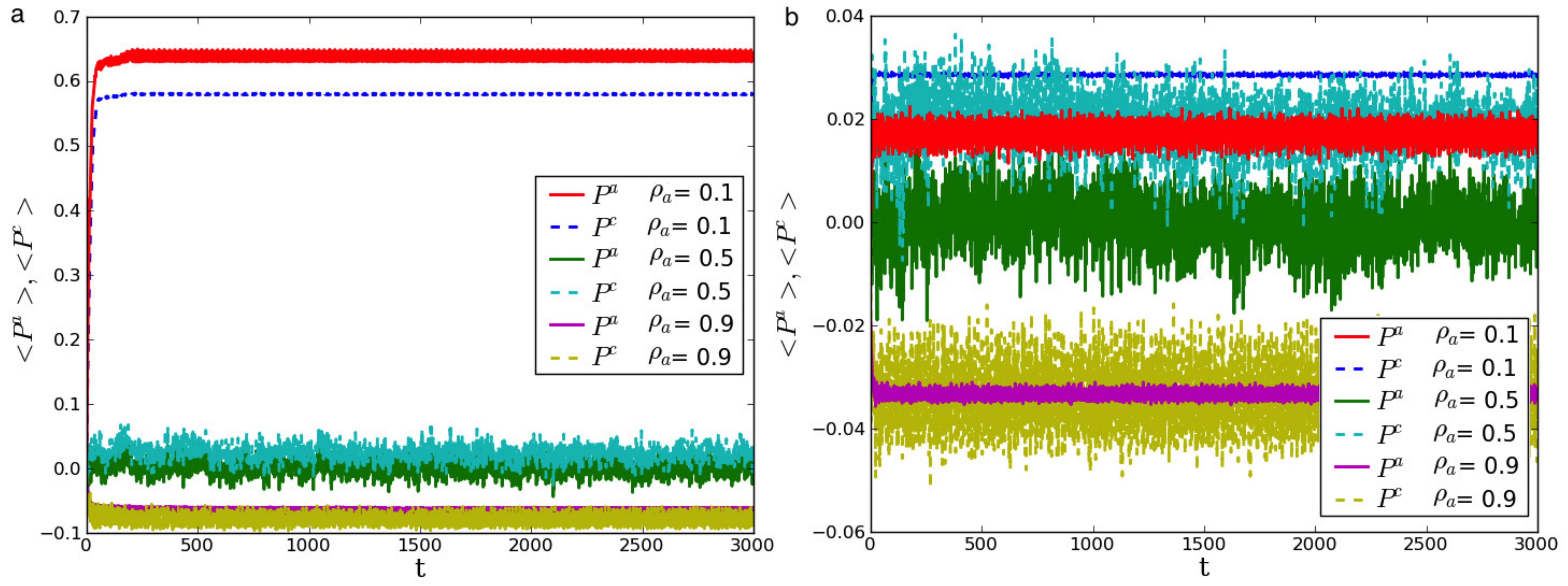}
\caption{Value of the average scores $P^{c}$ and $P^{a}$ (normalized over the relative number of conformist and nonconformist agents) over time, varying the value of $\rho_{a}$, i.e., the density of conformist agents. On the left, results achieved in scale-free networks. On the right results achieved in small-world networks. Results are averaged over $20$ different realizations.}
\label{fig:relative_score_P}
\end{figure}

In order to compare the two different agents' behaviors, we use the scores defined above. In particular, the local score $P$ to evaluate the system under the hypothesis \textbf{a} (related to the nature of nonconformity), and the global scores $(P^{c}$, $P^{a})$ to evaluate the system under the hypothesis \textbf{b}, i.e., by considering a radical nonconformity.  
Hence, varying the amount of conformist agents in the population, and considering both complex topologies (i.e., scale-free and small-world networks), we start analyzing the value of $\langle P \rangle$ over time --see Figure~\ref{fig:score_P}.
In general, results show that, according to $\langle P \rangle$, it is more convenient to behave as a nonconformist agent, with the exception of the cases where $\rho_{a} \to 1$ (we recall that such information is not known by agents).
On the contrary, by analyzing the global scores $P^{c}$ and $P^{a}$, we observe that in scale-free networks is more convenient to behave as a conformist for low values and for high values of $\rho_{a}$, whereas is slightly more convenient to be a nonconformist for intermediate values of $\rho_{a}$. 
Instead, in small-world networks, it is more convenient to be a nonconformist, with the exception of intermediate values of $\rho_{a}$ --see Figure~\ref{fig:relative_score_P}.

Then, considering the global scores $(P^{c}, P^{a})$, it is worth to highlight the differences between these two network topologies.
\subsection{The Behavior as a Degree of Freedom}
\begin{figure}[!ht]
\centering
\includegraphics[width=5.5in]{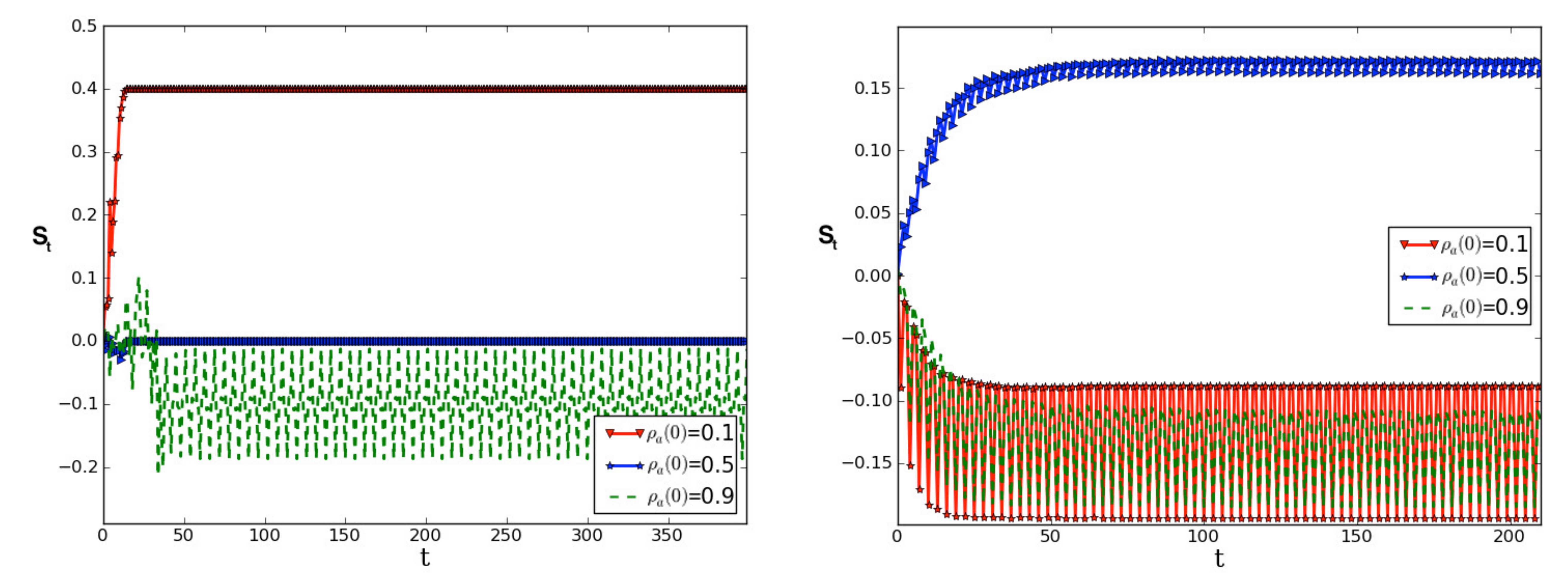}
\caption{Summation of states $S(t)$ over time, in the whole population of agents. Each curve refers to a different population containing, at $t=0$, a fraction $\rho_{a}$ of conformist agents (see the legend). On the left, results achieved in scale-free networks. On the right results achieved in small-world networks. Results are averaged over $20$ different realizations.}
\label{fig:state_S_adaptive}
\end{figure}

\begin{figure}[!ht]
\centering
\includegraphics[width=5.5in]{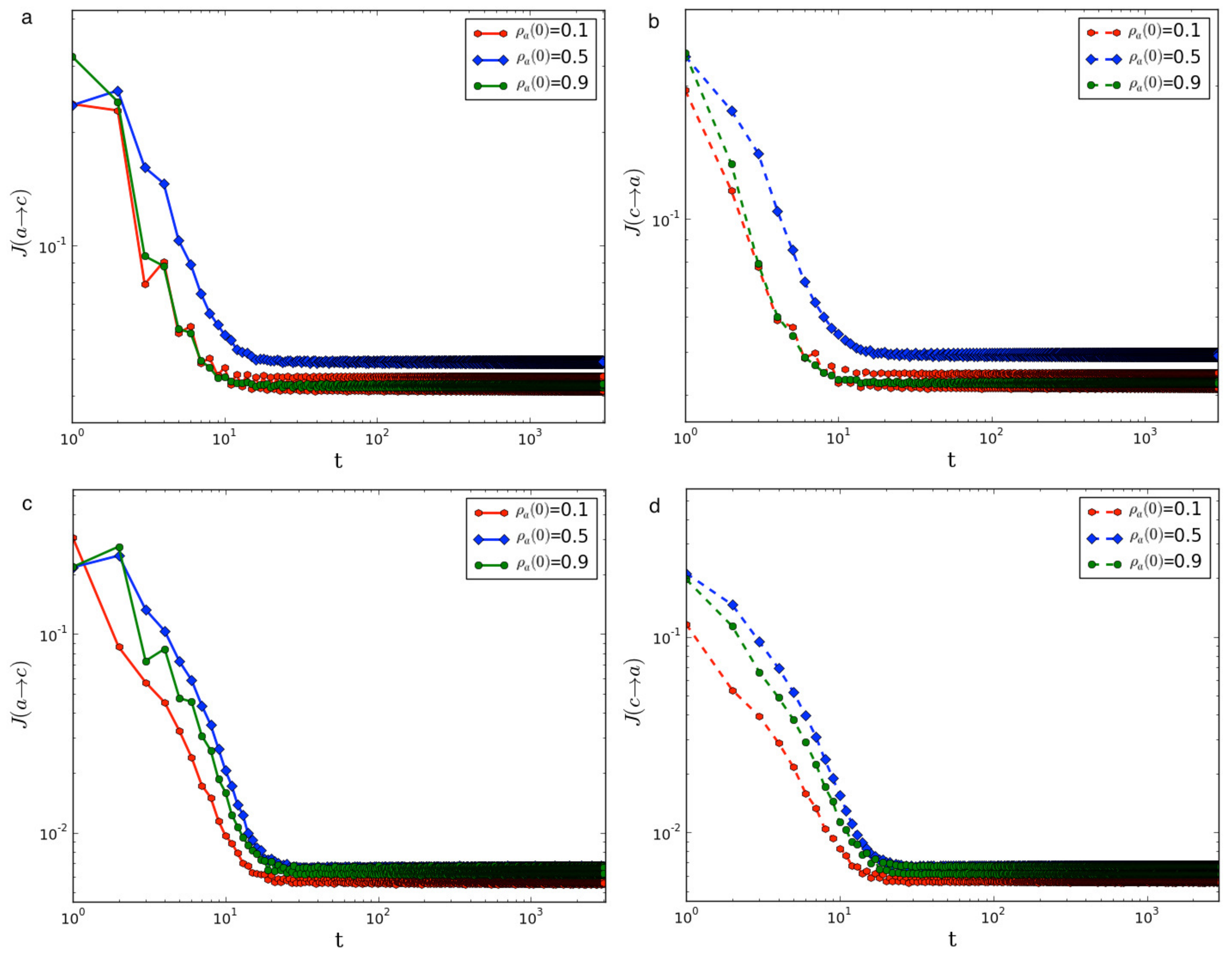}
\caption{Number of jumps $J$ from one behavior to another one (i.e., from conformist to nonconformist and vice versa) over time, in function of the initial density of conformist agents $\rho_{a}$. \textbf{a} Number of jumps from nonconformist to conformist in scale-free networks.  \textbf{b} Number of jumps from conformist to nonconformist in scale-free networks. \textbf{c} Number of jumps from nonconformist to conformist in small-world networks. \textbf{d} Number of jumps from conformist to nonconformist in small-world networks. Results are averaged over $20$ different realizations.}
\label{fig:jumps_adaptive}
\end{figure}

\begin{figure}[!ht]
\centering
\includegraphics[width=5.5in]{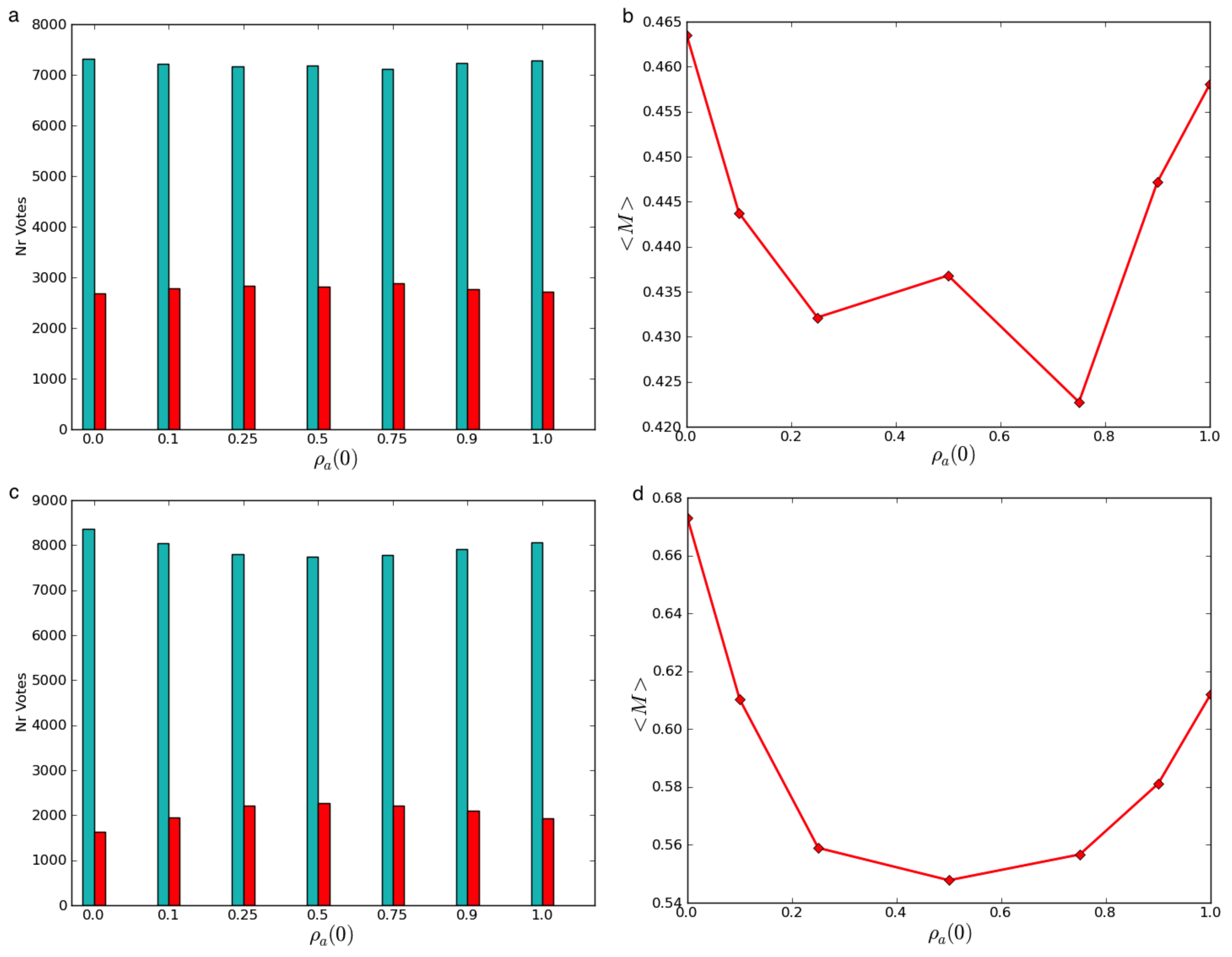}
\caption{\textbf{a} and \textbf{c} represent the number of agents in the two states, codified by two colors, in scale-free networks and in small-world networks, respectively. In particular, the cyan represents the number of agents having the predominant opinion, whereas the red that of agents with the other opinion. \textbf{b} and \textbf{d} represent the average magnetization of the system, in function of $\rho_{a}$, in scale-free networks and in small-world networks, respectively. Results are averaged over $20$ different realizations.}
\label{fig:magnetization_adaptive}
\end{figure}

\begin{figure}[!ht]
\centering
\includegraphics[width=5.5in]{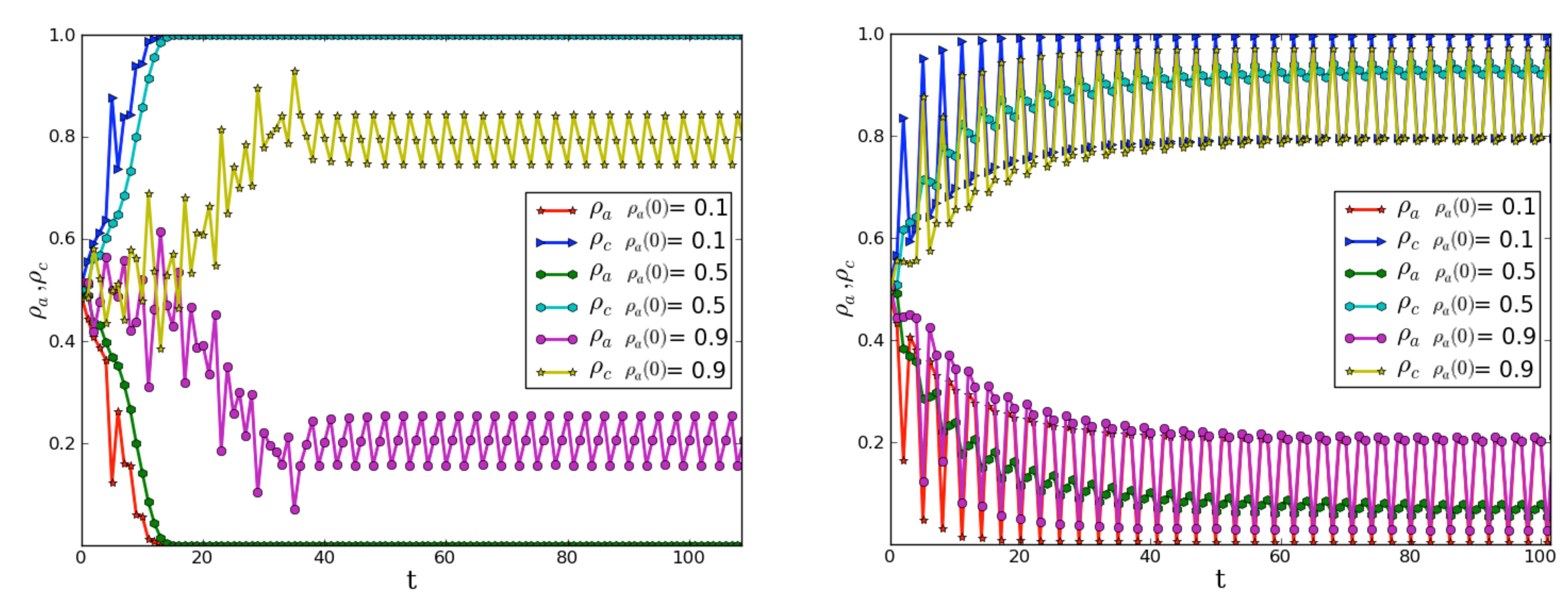}
\caption{Density of nonconformist agents (i.e., $\rho_{c}$) and of conformist agents (i.e., $\rho_{a}$) over time, varying the initial density $\rho_{a}(0)$. Jumps between the two behaviors are driven by the first hypothesis (i.e., ~\textit{a}), i.e., nonconformist agents behave as such only in relation to their social circle, but not in relation to the whole population. \textbf{a} Results achieved in scale-free networks. \textbf{a} Results achieved in small-world networks. Results are averaged over $20$ different realizations.}
\label{fig:density_absolute_adaptive}
\end{figure}

\begin{figure}[!ht]
\centering
\includegraphics[width=5.5in]{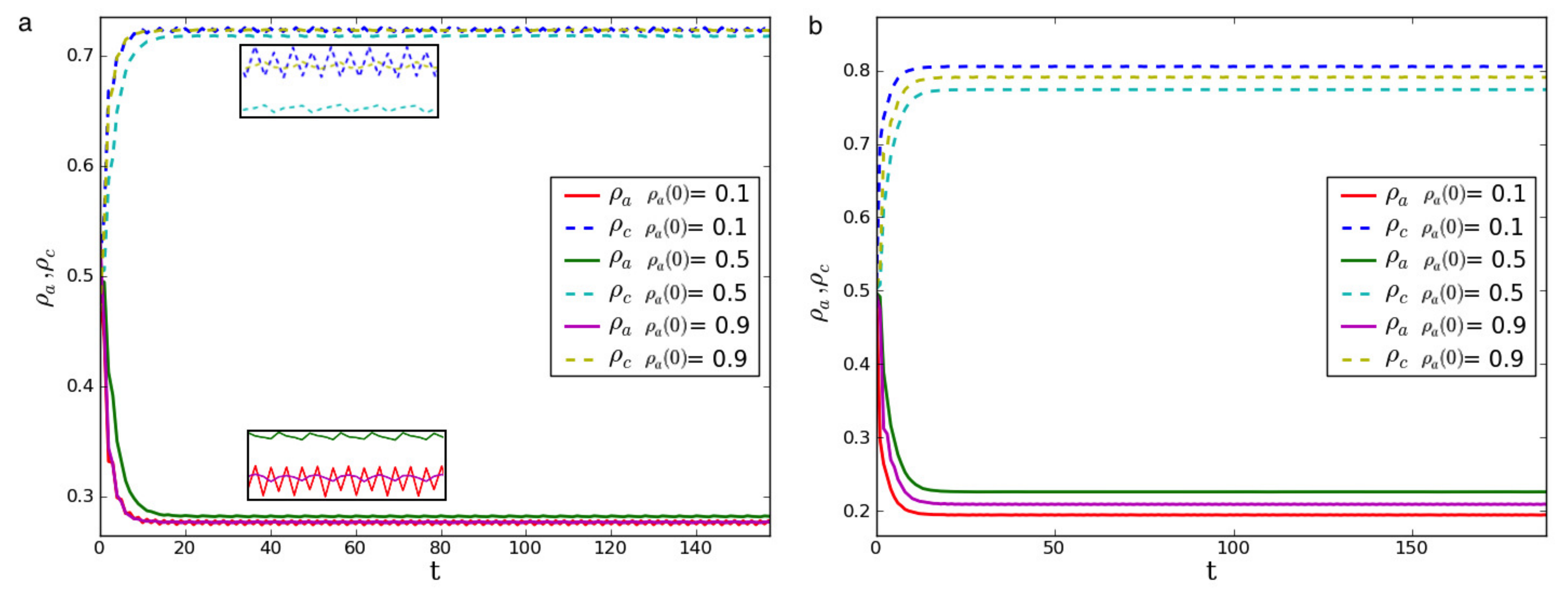}
\caption{Density of nonconformist agents (i.e., $\rho_{c}$) and of conformist agents (i.e., $\rho_{a}$) over time, varying the initial density $\rho_{a}(0)$. Jumps between the two behaviors are driven by the second hypothesis (i.e., ~\textit{b}), i.e., nonconformist agents behave as such in relation to the whole population. \textbf{a} Results achieved in scale-free networks. \textbf{a} Results achieved in small-world networks. Results are averaged over $20$ different realizations.}
\label{fig:density_relative_adaptive}
\end{figure}

Now, we analyze the proposed model considering the behavior as a degree of freedom, i.e., conformist agents can become nonconformist ones, and vice versa, over time. Similar mechanisms have already been investigated in~\cite{martins01,martins02}.
In this analysis, agents have been arranged in scale-free networks and in small-world networks, therefore we consider only complex topologies of the agent network.
The variation of $S(t)$, i.e., the summation of states in the whole population over time, is shown in Figure~\ref{fig:state_S_adaptive}.
It is interesting to note that, although we used two very different network topologies, the related outcomes are substantially very similar. 
In particular, the value of $S(t)$ is positive only for a low initial density of conformist agents (i.e., $\rho_{a}(0)$).
Since, in principle, each agent can modify its behavior many times during the evolution of the system, we analyze the number of changes (defined as jumps) from one behavior to another one over time, in the two kinds of complex networks --see Figure~\ref{fig:jumps_adaptive}.
On a quality level, once again, there are not strong differences between the two network topologies as, in each case, the number of jumps from one behavior to another one rapidly decreases in a few time steps ($\sim 10$). 
The number of jumps decreases, almost linearly in a double logarithmic scale, until it reaches a steady-state. The latter is characterized by the presence of a few agents that modify the behavior also after many time steps.
On the other hand, it is worth to highlight that the number of jumps $J$, once the steady-state is reached, is higher in scale-free networks than in small-world networks. Therefore, a quantitative difference can be found between the two network topologies.
Then, as shown in Figure~\ref{fig:magnetization_adaptive}, we analyze the average magnetization of the system.
The average magnetization $\langle M \rangle$ is lightly higher in small-world networks than in scale-free networks. Moreover, we note that, in both network topologies, $\langle M \rangle$ has higher values for extreme for values of $\rho_{a}(0)$ (i.e., lower than $0.25$ or higher than $0.75$), whereas it decreases as $\rho_{a}(0)$ has intermediate values (from $0.25$ to $0.75$).
Eventually, we study the transition between agents' behaviors. Results achieved by considering jumps driven by the hypothesis~\textit{a} are shown in Figure~\ref{fig:density_absolute_adaptive}.
In general, these results illustrate that agents find more convenient to behave as nonconformists, even if there are a lot of conformist agents at time $t=0$.
Figure~\ref{fig:density_relative_adaptive} shows results achieved by considering that jumps are driven by the hypothesis~\textit{b}.
We highlight that, by considering the second hypothesis, the densities of conformist and nonconformist agents, reached at the steady-state (after $\sim 10$ time steps), do not depend neither on the value of $\rho_{a}(0)$ nor on the topology of the agent network.
\subsection{Discussion}
We study a simple model of opinion dynamics with the aim to analyze the role of social influences. In particular, we consider the conformity, a behavior deemed relevant in these dynamics by social psychologists. 
In the proposed model, agents have a state that represents their opinion, and they are also endowed with an individual behavior, i.e., conformist or nonconformist. Moreover, we consider a system with only two opinions, hence agents have a state equal to $\pm 1$ that can vary over time.
In particular, at each time step, agents compute their state $s(t)$ in function of those of their neighbors, and also considering their individual behavior. Then, conformist agents modify their state according to the majority opinion of their social circle, whereas nonconformist agents do the opposite.
Since, we consider two levels of nonconformity, i.e., a local nonconformity (i.e., hypothesis \textbf{a}) and a global nonconformity (i.e., hypothesis \textbf{b}), we define two different kinds of score: the local score $P$ (related to the hypothesis \textbf{a}) and the global scores $(P^{c}, P^{a})$ (related to the hypothesis \textbf{b}). 
We recall that under the hypothesis \textbf{a}, the nonconformity has a local dimension as it is related only to the social circle of each agent, whereas under the hypothesis \textbf{b} the conformity has a global dimension as it is related to the whole population.
Hence, during the evolution of the system, agents aim to maximize their scores. 
The local score (i.e., $P$) is computed by comparing the state $s(t)$ of each agent with summation of states in the population $S(t)$. In particular, in the event an agent has a state equal to that of the majority of the population, its $P$ increases by $+1$, otherwise it decreases (by $-1$). The value of $P$ does not change only if $S(t) = 0$.
Instead, the global scores $(P^{c}, P^{a})$ are computed in function of the summation of states $S(t)$, but considering a more radical nonconformist behavior (hypothesis \textbf{b}). In particular, a nonconformist agent increases its score $P^{c}$ in the event its state $s(t)$ is opposite to $S(t)$, and it decreases when $s(t)$ is in accordance with $S(t)$; with the same logic, a conformist agent increases its score $P^{a}$ in the event its state $s(t)$ is in accordance with $S(t)$, and it decreases in the opposite case. 
As for the local score, also the global scores do not change if $S(t)=0$.
Global scores allow to evaluate the convenience of behaving as a conformist agent or not, from a more radical perspective.
A first analysis of the proposed model has been performed on fully-connected agent networks. The main result of numerical simulations, under this simple configuration, is that the density of conformist agents strongly affects the outcomes of the proposed model.
In particular, it is interesting to observe that as $\rho_{a}$ increases the system has a more stable behavior over time, i.e., the standard deviation of $S(t)$ falls to zero since first time steps (see left panel of Figure~\ref{fig:sum_states_fc}).
Then, further investigations have been performed by arranging agents in more complex topologies, as scale-free and small-world networks. 
Observing the variation of $S(t)$ (i.e., the summation of states in the population) over time, it is worth to highlight that the two different topologies of the agent network yield similar outcomes. In particular, the value of $S(t)$ reaches, in both cases, a steady-state characterized by small fluctuations around a small average value (see figure~\ref{fig:sum_states_complex}). This latter is usually positive, with the exception of systems having a high density of conformist agents in the population.
The variation of $S(t)$ over time shows a critical behavior, during first time steps, in scale-free agent networks when the density of conformist agents is very low ($<\ 0.25$). In particular, we observe a rapid increase of $S(t)$ followed by a decrease, until the steady-state is reached. 
We hypothesize that this behavior, not observed in small-world networks, is due to the presence of a few hubs (i.e., nodes with a very high degree which characterize the scale-free structure) that, at the beginning, strongly affect a great fraction of nodes. In particular, recalling that this phenomenon is achieved for a low density of conformist agents, it is more likely that many hubs be nonconformist agents. 
It is worth to observe that we found a great difference in the behavior of the system between fully-connected networks and complex networks. In particular, as discussed before, as $\rho_{a}$ increases in fully-connected networks as the system is more stable, whereas in both scale-free and small-world networks, after a steady-state is reached, the system seems more stable with low values of $\rho_{a}$.
On the other hand, by considering the average magnetization of the system, we found a difference between the two topologies of the agent network. In particular, the value of $\langle M \rangle$ is smaller in small-word networks than in scale-free networks. This result indicates that the two opinions tend to be distributed more uniformly in small-world networks.
The analysis of the scores provides further information about the proposed model. In particular, as shown in figure~\ref{fig:score_P}, agents achieve a higher local score when they behave as nonconformists, with the exception of populations characterized by the presence of many conformist agents.
Considering the global scores $(P^{c}, P^{a})$ we found that, when agents are arranged on scale-free networks, the conformist behavior is more convenient for low and high values of $\rho_{a}$, whereas the nonconformist behavior is slightly more convenient for intermediate values of $\rho_{a}$. 
On the other hand, in small-world networks it is more convenient to behave as a nonconformist agent, with the exception of intermediate values of $\rho_{a}$.
Therefore, in scale-free networks conformist agents often have a different state from that of the majority of the population, whereas this circumstance is more sporadic in small-world networks.
Finally, we analyzed the system allowing agents to change both their opinion and their behavior. In particular, at each time step each agent decides whether to behave as conformist or as a nonconformist. 
It is interesting to note that, by this new condition (i.e., the variation of behavior over time), the values of the average magnetization of the system are different from those achieved before. 
In particular, it is interesting to observe that $\langle M \rangle$ is now higher in small-world networks than in scale-free networks and, moreover, the values of $\langle M \rangle$ are higher by letting agents change behavior than in the opposite case (i.e., with unchangeable behaviors).
Furthermore, we define two different rules to let agents decide which behavior is more appropriate to be adopted. 
The first rule is based on the hypothesis \textit{a} about the nonconformist behavior, whereas the second rule is defined according to the hypothesis \textit{b}.
In general, results achieved by the utilization of the first rule indicate that, after a few time steps, almost all agents behave as nonconformist ones in both kinds of the agent network. The number of changes from one behavior to another one, shown in Figure~\ref{fig:jumps_adaptive}, confirms that a great fraction of agents decides its behavior during first time steps, whereas later only a few agents still perform jumps (we recall that by the term jump we mean change of behavior).
In the event the density of conformist agents is very high (i.e., $\> 0.75$), in scale-free networks, a small fraction of agents behave as conformist over time.
On the other hand, considering the second rule to drive the jumps, we observe that a small fraction of conformist agents is always present regardless of the initial density of conformist agents or the topology of the agent network.
Therefore, a few conformist agents can always survive in the system only if their nonconformist behavior is radical, then related to the whole population (i.e., hypothesis \textbf{b}).
As recalled in the introduction, the studying of the role of nonconformity has been already performed by other authors. In particular, in the work~\cite{weron01}, although this problem has been analyzed in a different way, as conclusion authors state that nonconformity plays a significant role in opinion dynamics, in a full accordance with our observations.
It is important to note that both in~\cite{weron01} and in~\cite{weron02}, authors identify two different kinds of nonconformity, i.e., anticonformity and independence; in the proposed model we consider the role of nonconformity in terms of anticonformity. 
Furthermore, we identify two different levels of nonconformity, one related to the social circle of each agent (i.e., a local nonconformity) and one related to the whole population (i.e., a global nonconformity).
\section{Conclusions}\label{sec:conclusions}
In this work, we study the role of social influences in opinion dynamics focusing our attention on the conformity, an important social behavior described by social psychologists. We propose an agent-based model, where a population is composed of conformist and nonconformist agents. Each agent, according to its individual behavior (i.e., conformist or nonconformist), defines its opinion in function of to those of its neighbors. In particular, conformist agents define their opinion according to that of the majority of their neighbors; instead, nonconformist agents do the opposite, i.e., they prefer to adopt the contrarian opinion. 
We introduce two different levels of nonconformity: one related to the social circle of each agent and one related to the whole population. The difference is that, in the former case, although nonconformist agents assume the contrarian opinion of their neighbors, they prefer to have the same opinion of the majority of agents in the whole population. Hence, the nonconformist behavior has only a local dimension. 
Instead, in the second case, nonconformist agents assume the contrarian opinion of their neighbors, but they prefer also to have an opinion different from that of the majority of the population. In doing so, the nonconformist behavior is more radical as it is not limited to the social circle of each agent (i.e., it has a global dimension). These two different levels of nonconformity require two different tools to analyze the evolution of the system. 
In particular, we introduce two kinds of scores that each agent aims to increase over time, i.e., the local score and the global scores. The local score considers the nonconformity only in relation to the social circle of each agent, whereas the global scores are related to the nonconformity extended to the whole population. Hence, the name assigned to the scores, i.e., local and global, are in accordance with the two levels of nonconformity which they refer to. Results of numerical simulations clearly show that the conformity is an important character that affects these dynamics (as it happens in real scenarios). 
Observing the evolution of the system in terms of summation of states, in all cases (i.e., varying the density of conformist agents) a steady-state is reached. The most interesting difference between the implemented complex topologies of the agent network is that, considering the global scores, often it is better to behave as a conformist agent in scale-free networks, whereas often it is better to behave as a nonconformist agent in small-world networks. 
On the other hand, considering the local score it is almost always more convenient to behave as a nonconformist agent. Furthermore, in the case agents can change also their behavior over time, a greater fraction of agents always prefers to behave as nonconformist regardless of the topology of the agent network. 
Therefore, from this point of view, it is important to highlight that the network topology does not always has a prominent role. On the other hand, we found a great difference when comparing the behavior of the proposed model between fully-connected networks and complex networks. 
In particular, we observed that in fully-connected networks conformist agents have the role of stabilizers, i.e., as their amount increases the system is more stable from the first time steps. Instead, in complex networks the opposite happens, i.e., in the presence of small fractions of conformist agents, after the steady-state is reached, the system is more stable than by introducing more conformists. 
Finally, from a socio-psychological perspective, the results achieved by the proposed model confirm the importance of the conformity. Moreover, since we found some differences among the outcomes achieved by introducing the two levels of conformity (i.e., local or related to the social circle, and global or related to the whole population), we deem useful that these differences be investigated also in real scenarios as they could lead to new interesting results also in social sciences.

\section*{Acknowledgments}
The author wishes to thank Serge Galam for his useful suggestions. This work was supported by Fondazione Banco di Sardegna.


\begin{thebibliography}{10}

\bibitem{loreto01}
Castellano, C. and Fortunato, S. and Loreto, V.:
Statistical physics of social dynamics.
\emph{Rev. Mod. Phys.} \textbf{81-2} 591--646 (2009)

\bibitem{sznajd01}
Sznajd-Weron, K. and Sznajd, J.:
Opinion Evolution in Closed Community.
\emph{International Journal of Modern Physics C} \textbf{11-6} 1157 (2000)

\bibitem{galam01}
Galam, S.:
Heterogeneous beliefs, segregation, and extremism in the making of public opinions.
\emph{Phys. Rev. E} \textbf{71-4} 046123 (2005)

\bibitem{galam02}
Galam, S.:
Local dynamics vs. social mechanisms: A unifying frame.
\emph{Europhys. Lett.} \textbf{70-6} 705-711 (2005)

\bibitem{galam03}
Galam, S.:
Sociophysics: a review of Galam models.
\emph{International Journal of Modern Physics C} \textbf{19-3} 409-440 (2008)

\bibitem{galam04}
Galam, S.:
Social paradoxes of majority rule voting and renormalization group.
\emph{Journal of Statistical Physics} \textbf{61} (1990)

\bibitem{krapviski01}
Krapivsky, P. L. and Redner, S.:
Dynamics of Majority Rule in Two-State Interacting Spin Systems.
\emph{Phys. Rev. Lett.} \textbf{102-1} 238701 (2003)

\bibitem{holme01}
Holme, P. and Newman, M.E.J.:
Nonequilibrium phase transition in the coevolution of networks and opinions.
\emph{Phys. Rev. E} \textbf{74-5} 056108 (2006)

\bibitem{bianconi01}
Halu, A. and Zhao, K. and Baronchelli, A. and Bianconi, G.:
Connect and win: The role of social networks in political elections.
\emph{Europhysics Letters} \textbf{90-23} 16002 (2013)

\bibitem{redner01}
Sood, V. and Redner, S.:
Voter Model on Heterogeneous Graphs.
\emph{Phys. Rev. Lett.} \textbf{94-17} 178701 (2005)

\bibitem{miguel01}
San Miguel, M. and Eguiluz, V.M. and Toral, R.:
Binary and Multivariate Stochastic Models of Consensus Formation.
\emph{Computing in Science and Engineering} \textbf{7-6} 67--73 (2005)

\bibitem{miguel02}
Castello, X. and Eguiluz, V.M. and San Miguel, M.:
Ordering dynamics with two non-excluding options: bilingualism in language competition.
\emph{New Journal of Physics} \textbf{8} (2006)

\bibitem{eger01}
Eger, S.:
Opinion dynamics under opposition.
http://arxiv.org/abs/1306.3134 (2013)

\bibitem{rapisarda01}
Pluchino, A., Rapisarda, A. and Garofalo, C.:
The Peter principle revisited: A computational study.
\emph{Physics A: Statistical Mechanics and its Applications} \textbf{389-3} 467-472 (2010)

\bibitem{rapisarda02}
Pluchino, A., Rapisarda, A. and Garofalo, C.:
Efficient promotion strategies in hierarchical organizations.
\emph{Physics A: Statistical Mechanics and its Applications} \textbf{20-1} 3496-3511 (2011)

\bibitem{williams01}
Fetta, A.G., Harper, P.R., Knight, V.A., Vieira, I.T. and Williams, G.E.:
On the Peter Principle: An agent based investigation into the consequential effects of social networks and behavioural factors.
\emph{Physics A: Statistical Mechanics and its Applications} \textbf{9-1} 2898-2910 (2012)

\bibitem{aronson01}
Aronson, E., Wilson, T.D. and Akert, R.M.:
Social Psychology.
Pearson Ed, (2006)

\bibitem{galam05}
Galam, S.:
From 2000 Bush–Gore to 2006 Italian elections: voting at fifty-fifty and the contrarian effect.
\emph{Quality \& Quantity} \textbf{41-4} 579-589 (2007)

\bibitem{weron01}
Nyczka, P., Sznajd-Weron, K., Cislo, J.:
Phase transitions in the q-voter model with two types of stochastic driving.
\emph{Physical Review E} \textbf{86} 011105 (2012)

\bibitem{weron02}
Nyczka, P., Sznajd-Weron, K.:
Anticonformity or Independence? - Insights from Statistical Physics.
\emph{Journal of Statistical Physics} \textbf{151} 174--202 (2013)

\bibitem{buechel01}
Buechel, B., Hellmann, T., and Klobner, S:
Opinion dynamics under conformity.
Institute of Mathematical Economics - Bielefeld- working paper 469 (2012)

\bibitem{javarone01}
Javarone, M.A., and Armano, G.:
Perception of similarity: a model for social network dynamics.
\emph{J. Phys. A: Math. Theor.} \textbf{46} 455102 (2013)

\bibitem{barabasi01}
Albert R. and Barabasi, A.L.:
Emergence of Scaling in Random Networks.
\emph{Science} \textbf{286} (5439) 509--512 (1999)

\bibitem{watts01}
Watts, D. J. and Strogatz, S. H.:
Collective dynamics of ``small-world'' networks.
\emph{Nature} 440-442 (1998)

\bibitem{mobilia01}
Mobilia, M. and Redner, S.:
Majority versus minority dynamics: Phase transition in an interacting two-state spin system.
\emph{Phys. Rev. E} \textbf{68-4} 046106 (2003)

\bibitem{martins01}
Martins, A. C. R. and Galam, S.:
Building up of individual inflexibility in opinion dynamics.
\emph{Phys. Rev. E} \textbf{87-4} 042807 (2013)

\bibitem{martins02}
Martins, A. C. R.:
Continuous opinions and discrete actions in opinion dynamics problems.
\emph{Int. J. Mod. Phys. C} \textbf{19} 617 (2008)

\end{thebibliography}
\end{document}